\documentclass[aps]{revtex4}
 \usepackage{amssymb} \usepackage{graphicx}
\usepackage{dcolumn} \usepackage{amsmath}

\begin{document}
\title{First-order symmetries of Dirac equation in curved background: a unified dynamical
symmetry condition} \thanks{Dedicated to Professor T. Dereli on the
occasion of his 60th birthday}
\author{\"{O}. A\c{c}{\i}k $^{1}$}
 \email{ozacik@science.ankara.edu.tr}
 \author{\"{U}. Ertem $^{1}$}
 \email{uertem@science.ankara.edu.tr}
\author{M. \"{O}nder $^{2}$}
 \email{onder@hacettepe.edu.tr}
 \author{A. Ver\c{c}in $^{1}$}
 \email{vercin@science.ankara.edu.tr}
\address{$^{1}$ Department of Physics, Ankara University, Faculty of Sciences,
06100, Tando\u gan-Ankara, Turkey\\
$^{2}$ Department of Physics Engineering, Hacettepe University,
06800, Beytepe-Ankara, Turkey.}

\date{\today}

\begin{abstract}

It has been shown that, for all dimensions and signatures, the most
general first-order linear symmetry operators for the Dirac equation
including interaction with Maxwell field in curved background are
given in terms of Killing-Yano (KY) forms. As a general gauge
invariant condition it is found that among all KY-forms of the
underlying (pseudo) Riemannian manifold, only those which Clifford
commute with the Maxwell field take part in the symmetry operator.
It is also proved that associated with each KY-form taking part in
the symmetry operator, one can define a quadratic function of
velocities which is a geodesic invariant as well as a constant of
motion for the classical trajectory. Some geometrical and physical
implications of the existence of KY-forms are also elucidated.
\end{abstract}


\maketitle

\section{INTRODUCTION}

In many evolutions taking place in a flat or curved background,
isometries of the underlying space-time metric lead to conservation
laws that also have clear geometrical meanings expressed by means of
their local generators, Killing vector fields. As space-time
transformations, flows of these fields specify the conserved
quantities as their flow invariants. However, since the beginning of
the 1970s, it has been recognized that many interesting properties
of a given space-time are intimately related to hidden symmetries of
its metric, which make themselves manifest in higher rank tensorial
objects that also provide additional conservation laws
\cite{Penrose,Walker,Hughston}. The building blocks of the
mathematical structure behind these hidden symmetries and associated
conserved quantities are the symmetric Killing tensors, KY-forms,
conformal Killing tensors and conformal KY (CKY)-forms. Their
defining relations are natural generalizations of those of Killing
vector fields and of conformal Killing vector fields. Although these
higher rank objects can be completely determined by the metric
itself (see for instance \cite{ozumav,ozumav1}), they are not
directly related to isometries or conformal transformations. But
they comprise, by definition, their generators in the rank one
sector of their hierarchy. In most of the cases these additional
conserved quantities lead to complete integrability of the
considered problem. This important step was initiated with the works
of Penrose and his collaborators \cite{Penrose,Walker}. They have
shown that it is the existence of a second-rank symmetric Killing
tensor which can be written as the ``square'' of a KY 2-form that
leads to Carter's fourth constant of motion, which is responsible
for the complete integrability of the geodesic problem in the Kerr
geometry \cite{Carter}. For other research fields utilizing KY and
CKY-forms and for earlier references for these forms we shall refer
to \cite{Benn-Kress,Collinson,Frulov1,Frulov2} and the references
therein.

An important place where some or all of these tensorial objects
enter the analysis is the study of symmetry operators for the
Dirac-type equations describing the motion in curved background with
or without additional interactions. By now it has become a
well-established fact that while CKY-forms take part in symmetry
operators, via the $R$-commuting argument for the massless Dirac
equation, KY-forms are indispensable in constructing first order
symmetries of the massless as well as massive Dirac equation in a
curved space-time. It is a four-dimensional Lorentzian space-time
where most of applications have taken place. The first seminal
studies in this context were carried out by Carter, McLenaghan and
Kamran \cite{Carter1,Carter2,Kamran}. The results of these earlier
studies, obtained in four dimensions for the massive or massless
Dirac equation in the absence of electromagnetic interactions, were
recently extended by Benn and his collaborators, to an arbitrary
dimension and signature
\cite{Benn-Charlton,Benn-Kress1,Benn-Kress2}.

In the absence of additional interactions, all KY or CKY-forms of
the space-time take part in the symmetry operator without any extra
restriction. However, when interactions are included, some
additional conditions arise which restrict the possible forms that
can enter into the symmetry operator. To the best of our knowledge,
these restrictions for a four dimensional curved background were
found for the first time by McLenaghan and Spindel
\cite{McLenaghan}. In searching for the most general symmetry
operator commuting with the Dirac equation in the presence of an
electromagnetic field, they found that the symmetry operator can be
constructed from KY-forms of the underlying background, provided
that they separately fulfill some conditions involving the field
itself. These included some conditions found before by Carter and
McLenaghan \cite{Carter2}. Some of these conditions were also
obtained before by Hughston {\it et al} in a slightly different
context: in searching quadratic first integrals for the charged
particle orbits in the charged Kerr background \cite{Hughston}.

In this study we first show that the main results of McLenaghan and
Spindel, that is the construction of the symmetry operators out of
KY-forms, can be extended to an arbitrary dimension and signature.
We then obtain a unified condition which allows one to specify which
KY-forms can take part in the symmetry operator, and hence define
hidden dynamical symmetry of the problem. This is an algebraic
condition and can be stated as follows; a KY-form of the curved
background enter the symmetry operator if and only if it Clifford
commutes with the force (Maxwell) field. In particular, we solve all
the consistency conditions and find a concise way to choose the
gauge in order to make  0-form component of the symmetry operator
constant. Owing to a non-integrated consistency condition, this
point has remained ambiguous in the literature.

Our results include the results of Benn and his collaborators when
the electromagnetic field is turned off. Finally, we prove that the
quadratic functions of velocities defined in terms of each KY
$p$-form entering the symmetry operator is not only a geodesic
invariant but also, for an arbitrary number of dimension and
signature, a constant of motion for the classical trajectory.

We shall mainly use the notation of \cite{Benn-Tucker} and adopt the
following conventions and terminology. The underlying base manifold
is supposed to be an $n-$dimensional pseudo-Riemannian manifold with
arbitrary signature. Covariant derivative of spinors, that is of
sections of a bundle carrying an irreducible representation of the
(real or complexified) Clifford algebra, with respect to the vector
field $X$ is denoted by $S_{X}$. Then the Dirac operator on spinors
is $\displaystyle{\not}S=e^{a}S_{X_{a}}$, where the local co-frame
$\{e^{a}\}$ is the dual to the tangent frame $\{X_{a}\}$. Summation
convention over repeated indices will be used throughout the paper.
Juxtaposing $e^{a}$ and $S_{X_{a}}$, or any other operator, or form
will denote the Clifford multiplication. When acting on forms
$S_{X}$ and $\displaystyle{\not}S$ will be denoted, in terms of the
pseudo-Riemannian connection $\nabla$, as $\nabla_{X}$ and
\begin{eqnarray}
\displaystyle{\not}d=e^{a}\nabla_{X_{a}}=d-\delta\;.\nonumber
\end{eqnarray}
Here $d=e^{a}\wedge\nabla_{X_{a}}$ and
$\delta=-i_{X_{a}}\nabla_{X^{a}}$ denote the exterior derivative and
the co-derivative written in terms of covariant derivative operator
$\nabla_{X}$. $\wedge$ is the exterior product and $i_{X}$ will
represent the interior derivative with respect to $X$ whose action
on an arbitrary $p$-form $\alpha$ is defined, for all vector fields
$Y_{j}$, by
\begin{eqnarray}
(i_{X}\alpha)(Y_1,\dots,Y_{p-1})=p
\alpha(X,Y_1,\dots,Y_{p-1})\;.\nonumber
\end{eqnarray}
For dual basis elements we have
$i_{X_{b}}e^{a}=e^{a}(X_{b})=\delta^{a}_{b}$.

The rest of the paper is organized as follows. In section II, the
general form of the first order symmetry operator of the Dirac
equation with a potential term is specified. This is achieved by
constructing and then by solving all consistency equations except
the equation for the $0$-form component of the non-derivative term
of the symmetry operator. The unified dynamical symmetry condition
announced in the title is established in section III by analyzing
higher degree components of the consistency equations. Special cases
of this condition and the integration of the remaining $0$-form
component are also presented there. In section IV, the
correspondence between KY and closed CKY-forms are studied and Yano
vectors are introduced. Implications of the existence of Yano
vectors related to symmetry analysis and to the global structure of
the underlying space-time are given in the same section. In section
V, the first integrals of geodesic equations, constants of motion of
classical trajectories and their connection with the KY-forms and
the mentioned dynamical symmetry condition are considered.
Derivation of consistency equations and the contraction of curvature
$2$-forms with a Yano vector are given in two appendixes. In the
Appendix B determination of upper bounds for the numbers of linearly
independent KY-forms is also provided. Section VI concludes the
paper.

\section{FIRST ORDER SYMMETRY OPERATORS OF THE DIRAC EQUATION}

We set out to our analysis by considering the Dirac equation
\begin{eqnarray}
(\displaystyle{\not}S+{\rm{i}}A)\psi=m\psi
\end{eqnarray}
for a complex spinor field $\psi$ and propose the following first
order linear symmetry operator
\begin{eqnarray}
L=2\omega^{a}S_{X_a}+\Omega\;,
\end{eqnarray}
such that $L$ Clifford commutes with
$\displaystyle{\not}S+{\rm{i}}A$. Equation (1) describes the motion
of a massive, charged and spin $1/2$ particle, with unit charge and
mass $m$, interacting with the curved background encoded in the
Dirac operator $\displaystyle{\not}S$ and force field represented by
the potential form $A$.  By fixing the charge from the outset, we
assume the effective coupling to the spinor field. In the beginning
$A$ is allowed to be an arbitrary inhomogeneous form, and the
consistency conditions are derived in this general context. Later
on, $A$ will be taken to be a 1-form. In that case, the term
involving $A$ in equation (1) describes coupling to the Maxwell
field $F=dA$. $A$ may also involve potential terms of
non-electromagnetic origin such as those of conservative forces. In
the latter case, $F$ will be referred to as a force field. In
equation (2) $\omega^{a}$ and $\Omega$ are
$\mathbb{Z}_{2}$-homogeneous, both even or both odd forms which act,
like $A$, on spinor fields by the Clifford multiplication.

In recent studies \cite{Benn-Charlton,Benn-Kress1,Benn-Kress2}  it
has been proved that use of the graded Clifford commutator $[,]$
considerably eases the calculations in analyzing the symmetries of
the Dirac-type operators. For a $p$-form $\alpha$ and an
inhomogeneous Clifford form $\beta=\sum_{q}\beta_{(q)}$, this
commutator is  defined as
\begin{eqnarray}
[\alpha,\beta]=\alpha\beta-\sum_{q}(-1)^{pq}\beta_{(q)}\alpha
\;.\nonumber
\end{eqnarray}
If $\alpha$ is a one-form this transforms, in terms of the main
involution $\eta$ (which leaves the even forms invariant and changes
the sign of the odd forms) of the Clifford algebra, to
$[\alpha,\beta]=\alpha\beta-\beta^{\eta}\alpha$. Let us suppose that
\begin{eqnarray}
[\displaystyle{\not}S, L]+{\rm{i}}[A, L]=0\;,
\end{eqnarray}
is satisfied and let us call $L$ even (odd)  when $\omega^{a}$ and
$\Omega$ are both even (both odd). When $L$ is even its graded
commutator with the Dirac equation becomes the usual Clifford
commutator $[,]_{-}$ and irrespective of $n,\;L$ is a symmetry
operator. That is, it Clifford commutes with the  Dirac equation and
maps a solution to another. When $L$ is odd it anti-commutes with
Dirac equation and fails to be a symmetry operator. However, in such
a case if $n$ is even $Lz$ is a symmetry operator since the volume
form $z$ anti-commutes with odd forms. (In such a case $A$ must be
an odd form.)

\subsection{The Main Consistency Equations}

To determine $\omega^{a}$'s and $\Omega$ in equation (2), we equate
the symmetrized coefficients of the covariant derivatives of each
order to zero in equation (3). The second order derivatives come
only from the first bracket of (3) such that
\begin{eqnarray}
 [\displaystyle{\not}S, L]=2i_{X_{b}}\omega^{a}[S^{2}(X_{a},X^{b})+
 S^{2}(X^{b},X_{a})]+\cdots\;,
\end{eqnarray}
where $i_{X}$ is the interior derivative,
\begin{eqnarray}
S^{2}(X,Y)=S_{X}S_{Y}-S_{\nabla_{X}Y} \;,\nonumber
\end{eqnarray}
denotes the second covariant derivative and the ellipsis stands for
lower order terms. Equating the coefficients of the second order
derivatives to zero in equation (3) yields
\begin{eqnarray}
 i_{X_b}\omega^a+i_{X^{a}}\omega_{b}=0\;,
\end{eqnarray}
for all $a,b=1,2,\dots,n$. This is satisfied if and only if
$\omega^a=i_{X^a}\omega$ where
 $\omega $ is possibly a $\mathbb{Z}$-inhomogeneous form. It has been
shown in Appendix A that, in view of (5), by equating the
coefficients of equal power of derivatives in equation (3) we obtain
\begin{eqnarray}
\nabla_{X^{a}}\omega &=& i_{X^{a}}\varphi-{\rm{i}}[A, i_{X^{a}}\omega]\;,\\
\nabla^{2}(X_{a},X^{a})\omega&=&\displaystyle{\not}d\varphi+2{\rm{i}}
(i_{X^{a}}\omega)^{\eta}\nabla_{X_{a}}A-{\rm{i}}[A,\Omega]\;,
 \end{eqnarray}
where $\varphi=\displaystyle{\not}d \omega-\Omega$. These constitute
two main sets of the consistency conditions that specify the form of
possible symmetry operators of the Dirac equation, for all
dimensions and signatures, in which $A$ is a general form.
Henceforth, we take $A$ to be a 1-form and write, for an arbitrary
form $\alpha$
\begin{eqnarray}
[A, \alpha]=2i_{\tilde{A}}\alpha\;.\nonumber
\end{eqnarray}
Here $\tilde{A}$ is the metric dual of $A$, which in terms of the
metric $g$ of the background is defined by $A(X)=g(\tilde{A},X)$ for
all $X$.

\subsection{Solutions: Emergence of KY-Forms}

Let us first concentrate on equation (6), which for 1-form $A$ can
be rewritten as
 \begin{eqnarray}
 \nabla_{X^{a}}\omega = i_{X^{a}}(\varphi+2{\rm{i}}i_{\tilde{A}}\omega) \;.
 \end{eqnarray}
Applying $i_{X_{a}}$ to both sides of this equation, we firstly see
that $\delta \omega=0$, that is, $\omega$ must be co-closed. On the
other hand, by applying $e_{a}\wedge$ we obtain
 \begin{eqnarray}
 d\omega = \pi (\varphi+2{\rm{i}}i_{\tilde{A}}\omega)\;,
 \end{eqnarray}
 where the linear map $\pi$ scales each form component by its degree :
 $\pi(\alpha)=e^{a}\wedge i_{X_{a}}\alpha$. The $(p+1)$-form
 component of equation (9) reads, for $p=0,1,\dots,n-2$, as
  \begin{eqnarray}
 \varphi_{(p+1)}=\frac{1}{p+1}d\omega_{(p)}-2{\rm{i}}i_{\tilde{A}}\omega_{(p+2)}\;,
 \end{eqnarray}
and for $p=n-1$, as $\varphi_{(n)}=n^{-1}d\omega_{(n-1)}$. In view
of the last two equations, equation (8) implies that each $p$-form
component of $\omega$ must obey
 \begin{eqnarray}
 \nabla_{X_{a}}\omega_{(p)} =\frac{1}{p+1} i_{X_{a}}d\omega_{(p)}\;,
 \end{eqnarray}
which is the well-known KY-equation.

From equation (10) and by the fact that $\omega$ is co-closed
 we also obtain
\begin{eqnarray}
\Omega_{(p+1)}&=&\frac{p}{p+1}d\omega_{(p)}+2{\rm{i}}i_{\tilde{A}}
\omega_{(p+2)}\;,\\
\Omega_{(n)}&=&(1-\frac{1}{n})d\omega_{(n-1)}\;,
\end{eqnarray}
for $p=0,1,\dots,n-2$. Note that for $p=0,1,\dots,n-1$ we have
\begin{eqnarray}
i_{\tilde{A}}\Omega_{(p+1)}=p\nabla_{\tilde{A}}\omega_{(p)}\;.
\end{eqnarray}
These provide the non-derivative term, except for its 0-form
component, of the symmetry operator in terms of KY-forms and the
potential.

We now turn to the implications of equation (7). Differentiating
equation (8) once more, we  obtain
\begin{eqnarray}
 \nabla^{2}(X_{a},X^{a})\omega=-\delta(\varphi+2{\rm{i}}i_{\tilde{A}}\omega)\;,
\end{eqnarray}
and by combining this with equation (7), we arrive at
\begin{eqnarray}
 i_{\tilde{A}}\Omega-\frac{{\rm{i}}}{2}d\Omega =
 \delta (i_{\tilde{A}}\omega)+(i_{X^{a}}\omega)^{\eta}\nabla_{X_{a}}A\;.
\end{eqnarray}
This can be used to obtain the 0-form component of $\Omega$ and
possible relations among its higher degree components. For this
purpose we apply the general relation
\begin{eqnarray}
[\delta, i_{X}]_{+}=-i_{X^{a}}i_{\nabla_{X_{a}}X}\;,\nonumber
\end{eqnarray}
to KY-forms and obtain
 \begin{eqnarray}
 \delta(i_{\tilde{A}}\omega)&=&-i_{X^{a}}i_{\nabla_{X_{a}}\tilde{A}}\omega\;.
 \end{eqnarray}
We now make use of
\begin{eqnarray}
  (i_{X^{a}}\omega)^{\eta}\kappa_{a}=
 \kappa_{a}\wedge i_{X^{a}}\omega-i_{\tilde{\kappa}_{a}}i_{X^{a}}\omega,
  \end{eqnarray}
where $\kappa_{a}$'s are 1-forms. This relation is a direct result
of the standard Clifford multiplication rule for a right
multiplication of an arbitrary form by a 1-form. Using (18), by
taking $\kappa_{a}=\nabla_{X_{a}}A$, and (17) in (16) we get
\begin{eqnarray}
 i_{\tilde{A}}\Omega-\frac{{\rm{i}}}{2}d\Omega =
 (\nabla_{X_{a}}A)\wedge i_{X^{a}}\omega\;.
  \end{eqnarray}
The $1$-form component of this relation reads as
\begin{eqnarray}
d\Omega_{(0)} &=&-2{\rm{i}}( i_{\tilde{A}}\Omega_{(2)}
 -i_{X^{a}}\omega_{(1)}\nabla_{X_{a}}A)\;,\nonumber\\
&=&-2{\rm{i}}( \nabla_{\tilde{A}}\omega_{(1)}
 -\nabla_{\tilde{\omega}_{(1)}}A)\;.
 \end{eqnarray}
In obtaining the second equality we made use of (14). The equation
(20) will be integrated in the next section after the higher degree
components of equation (19) are further analyzed to uncover the
symmetry condition stated in the title.

\section{A Unified Dynamical Symmetry Condition}

In this section we first show that equation (16) contains an
important algebraic condition which plays a prominent role in
deciding which KY-forms can take part in the symmetry operator. To
see this, we first recall that $\Omega=d\omega-\varphi$ which
implies
\begin{eqnarray}
d\Omega&=&-d\varphi=2{\rm{i}}di_{\tilde{A}}\omega\;,\nonumber\\
i_{\tilde{A}}\Omega &=&i_{\tilde{A}}d\omega-i_{\tilde{A}}\varphi=\pi
i_{\tilde{A}}\varphi\;.\nonumber
 \end{eqnarray}
In the last equality we made use of (9). On substituting these
relations into (16), we get
\begin{eqnarray}
\pi(i_{\tilde{A}}\varphi)=-\displaystyle{\not}d
i_{\tilde{A}}\omega+(i_{X^{a}}\omega)^{\eta}\nabla_{X_{a}}A\;.
\end{eqnarray}
We now make use of
\begin{eqnarray}
 [\displaystyle{\not}d, i_{X}]_{+}=\nabla_{X}+e^{a}i_{\nabla_{X_{a}}X}\;,
\end{eqnarray}
where $[,]_{+}$ denotes the Clifford anti-commutator. Using (18),
once again by taking $\kappa_{a}=\nabla_{X_{a}}A$, and (22) in the
equation (21) we arrive at
\begin{eqnarray}
 \pi(i_{\tilde{A}}\varphi)-i_{\tilde{A}}\displaystyle{\not}d\omega+
 \nabla_{\tilde{A}}\omega=-\frac{1}{2}[dA, \omega]_{-}\;,
 \end{eqnarray}
where we have also used the following relation:
\begin{eqnarray}
 (\kappa_{a}\wedge i_{X^{a}}-e^{a}\wedge i_{\tilde{\kappa}_{a}})\omega
 =-\frac{1}{2}[dA, \omega]_{-}\;.
\end{eqnarray}

It is now easy to verify that, by the KY-equation (11) and by the
equation (10), the left hand side of (23) vanishes and we obtain an
important condition $[dA, \omega]_{-}=0$. On the other hand, as is
apparent in equation (24), since the Clifford commutator of an
arbitrary form by a 2-form does not change the degrees of its
components, each p-form component of $\omega$ must Clifford commute
with the force field $F=dA$:
 \begin{eqnarray}
 [F, \omega_{(p)}]_{-}=0\;.
 \end{eqnarray}
This gauge invariant condition means that among the KY-forms
admitted by the underlying space-time, only those which satisfy the
above condition take part in the symmetry operators. Moreover, it
must be emphasized that it is the force-field, not the potential
form itself, that plays a selective role in this regard.

At this point  we should note that, in view of equations (10), (14)
and equation (21) derived above, the higher degree components of
equation (19) lead us again to condition (25) and yield nothing new.

In order to integrate equation (20) as well as to obtain more
practical statements resulting from the equation (25), we rewrite it
as
\begin{eqnarray}
i_{X^{a}}F\wedge i_{X_{a}}\omega_{(p)}=0\;.
 \end{eqnarray}
This is obviously satisfied for $p=0$ and also for $p=n$. Hence the
extreme cases $p=0,n$ do not impose any condition on $F$ and on
possible KY-forms that take part in the symmetry operator $L$. In
general, a KY $p$-form $\omega_{(p)}$ takes part in the symmetry
operator if and only if $F$ is in the kernel of the operator
$i_{X_{a}}\omega_{(p)}\wedge i_{X^{a}}$. More practical refinements
of this condition are attained for intermediate values of $p$. We
first should note that the 0-form component $\omega_{(0)}$ of a
KY-form can be any function and $\omega_{(1)}$ is the dual of a
Killing vector field. Moreover, $\omega_{(n)}$ is a constant
(parallel), that is, it is a constant multiple of the volume form:
$\omega_{(n)}=kz$. For $p=1$ we can write $\omega_{(1)}=\tilde{K}$,
where $K$ is a Killing vector field. In this case, equation (26)
reduces to
\begin{eqnarray}
i_{K}F=0\;,
\end{eqnarray}
which implies that $F$ remains invariant under the flow generated by
$K$. That is, ${\cal L}_K F=0$, where ${\cal L}_K$ denotes the Lie
derivative in the direction of $K$. In other words, even the
generator of the isometries will take part in the symmetry operators
if they fulfill condition (27).

In terms of the potential $1$-form $A$ the condition (27) reads as
$i_K(e^{a}\wedge \nabla_{X_a} A)=0$ which, in view of
$i_{\nabla_{X}K}A=i_{\tilde{A}}\nabla_{X}\omega_{(1)}$, can be
evaluated to obtain
\begin{eqnarray}
0=\nabla_{K}A-di_{K}A-\nabla_{\tilde{A}}\omega_{(1)}.\nonumber
\end{eqnarray}
By comparing this relation with equation (20) and then by
integrating we obtain
\begin{eqnarray}
\Omega_{0}=2{\rm{i}} i_{K}A\;,
\end{eqnarray}
up to a constant. This relation, which was not recognized in the
literature before, provides a concise way of choosing the gauge in
order to make $\Omega_{0}$ constant. In such a case, only duals of
Killing vector fields whose flows preserve the potential can appear
in the symmetry operator. An equivalent but more instructive way of
deriving (28) goes as follows. Recalling the fact that for a 1-form
$\beta$ the metric dual of $\nabla_{X}\beta$ is
$\nabla_{X}\tilde{\beta}$, we can rewrite condition (20) as
\begin{eqnarray}
\widetilde{d\Omega}_{(0)} =-2{\rm{i}}[\tilde{A},K]_{L}=2{\rm{i}}
{\cal L}_{K} \tilde{A}\;.\nonumber
\end{eqnarray}
Here $[,]_{L}$ denotes the usual Lie bracket of vector fields, and
we have also used the zero torsion condition. Since $K$ is a Killing
field, the above relation simply reads as $d\Omega_{(0)}=2{\rm{i}}
{\cal L}_{K} A$ and then the usual action of Lie derivative on
differential forms produces the desired relation in view of (27).

For a KY 2-form $\omega_{(2)}=2^{-1}\omega_{ab}e^{a}\wedge e^{b}$
the condition (26) reads as $\omega_{a[b}F_{c]}^{\;a}=0$ for all
values of $b$ and $c$ where the square bracket stands for the
anti-symmetrization of the enclosed indices and $F_{cb}$ are the
components of $F$ in the same $\{e^{a}\}$ basis. This condition is
the tensor-language version \cite{Carter2} of the condition first
obtained by Hughston {\it et al} \cite{Hughston}. Our condition
(25), or equivalently (26), are natural generalizations of this and
similar conditions. This reflects the efficiency of the Clifford
calculus in this context.

To discuss condition (26) for higher forms it is convenient to
rewrite it, in terms of $(p+1)$-forms defined by
\begin{eqnarray}
\beta_{a}=F\wedge i_{X_{a}}\omega_{(p)}\;,\nonumber
\end{eqnarray}
as $i_{X^{a}}\beta_{a}=0$. This is also satisfied for $p=0, n$. In
the latter case $\beta_{a}=0$ are identically satisfied, since all
$\beta_{a}$'s are $(n+1)$-forms. For $p=n-1$ we have
$\beta_{a}=f_{a}z$ for some set of functions $f_{a}$ such that
equation (26) amounts to $i_{V}z=0$ for $V=f_{a}X^{a}$. As the map
$\varphi_{z}$ defined by $\varphi_{z}(V)=i_{V}z$ between the vector
fields and $(n-1)$-forms is an isomorphism, $i_{V}z$ vanishes if and
only if $V=0$, that is if and only if all $f_{a}$ vanish. This is
equivalent to $F\wedge i_{X_{a}}\omega_{(n-1)}=0$ for all $X_{a}$,
or to
\begin{eqnarray}
i_{X_{a}}F\wedge \omega_{(n-1)}=0\;.
\end{eqnarray}
In the case of $p=n-2,\;\beta_{a}$'s are $(n-1)$-forms and each one
can be written, in terms of uniquely determined  $1$-form
$\sigma_{a}=\sigma_{ab}e^{b}$, as $\beta_{a}=^{\ast}\sigma_{a}$. In
such a case equation (26) amounts to
$\sigma_{a}\wedge\tilde{X}^{a}=0$, which implies that
$\sigma_{ab}=\sigma_{ba}$. This can succinctly be expressed as
 \begin{eqnarray}
F\wedge (\tilde{X}_{a}\wedge i_{X_{b}}-\tilde{X}_{b}\wedge
i_{X_{a}})\omega_{(n-2)}=0\;,\nonumber
\end{eqnarray}
for $a,b=1,\dots,n$. Note that the considered cases
$p=0,1,n-2,n-1,n$ exhaust all possible forms of equation (25) in
four dimensions. More appealing versions of some of these conditions
will appear in the next section (see equation (34) and the remarks
followed).

Finally in this section we should note that, when $A$ is zero we
recover exactly the results of Benn and Kress for all dimensions and
signatures, obtained in
\cite{Benn-Charlton,Benn-Kress1,Benn-Kress2}. In such a case, since
the condition (25) disappears, all KY-forms of the underlying
space-time take part in the symmetry operators and all of the
non-derivative terms explicitly given by equations (12) and (13),
except $\Omega_{0}$ which is constant, are exact.

\section{CORRESPONDENCE BETWEEN KY-FORMS AND CLOSED CKY-FORMS}

The appearance of higher rank KY-forms in the symmetry operator
indicates the presence of dynamical symmetries which are not
isometries. We shall now establish a general one-to-one
correspondence between the KY-forms and closed CKY-forms and, as a
particular case, we shall show that for each KY $(n-1)$-form there
exists a uniquely determined conformal transformation. This
transformation is generated by a locally gradient field whose
integral curves are pre-geodesics. In this way we shall see that the
higher KY-forms are related to special CKY-forms and, in the case of
$(n-1)$-forms, to the special conformal transformations rather than
isometry transformations.

For the two points mentioned above, let us consider the CKY-equation
\begin{eqnarray}
\nabla_{X}\rho_{(p)} =\frac{1}{p+1} i_{X}d\rho_{(p)}-\frac{1}{n-p+1}
\tilde{X}\wedge \delta \rho_{(p)}\;,
\end{eqnarray}
which has the well-established Hodge duality invariance and
conformal covariance. A p-form $\rho_{(p)}$ is called CKY p-form if
and only if it satisfies equation (30) for all vector fields $X$. It
is obvious that a $p$-form is a KY $p$-form if and only if it is a
co-closed CKY $p$-form. A less obvious fact is that a $(n-p)$-form
is a KY $(n-p)$-form if and only if it is the Hodge dual of a closed
CKY $p$-form. Indeed, the Hodge dual of (30) for closed CKY forms
yields
\begin{eqnarray}
\nabla_{X_{a}}\;^{\ast}\rho_{(p)} =\frac{1}{n-p+1} i_{{X}_{a}}d
^{\ast}\rho_{(p)}\;,\nonumber
\end{eqnarray}
where we have made use of the relation
$i_{X}\;^{\star}\phi=^{\star}(\phi\wedge\tilde{X})$ which can be
considered as the definition of the Hodge map. Since $\rho_{(p)}$ is
closed it is locally exact such that $\rho_{(p)}=d\alpha$, where the
$(p-1)$-form $\alpha$ may be termed as the {\it KY-potential} for
the KY $(n-p)$-form $\omega_{(n-p)}=^{\ast}d\alpha$. Note that
$0$-forms of KY and CKY coincide and that any CKY $n$-form is of the
form $\rho_{(n)}=fz$, where $f$ can be any differentiable function.
As a result, the Hodge map establishes a vector space isomorphism
between the vector space of all KY $p$-forms and that of all closed
CKY $(n-p)$-forms for all values of $p$.

\subsection{KY $(n-1)$-forms and Yano Vectors}

In the case of KY $(n-1)$-forms we can write
\begin{eqnarray}
\omega_{(n-1)}=i_{Y}z=^{\ast}\tilde{Y}\;,
\end{eqnarray}
where the vector field $Y$ is the metric dual of the associated
closed CKY 1-form $\tilde{Y}$. $Y$ will be referred to, following
McLenaghan and Spindel, as the {\it Yano vector}. In that case,
connection with the conformal transformations can be established in
a more instructive manner. As the space of $(n-1)$-forms and that of
$1$-forms are isomorphic, exactly as in (31), any form of the former
type can be written as the Hodge dual of a uniquely determined
latter type. Recalling that the divergence of a vector field $U$
with respect to the volume form $z$ is defined by
$L_{U}z=(div_{z}U)z$, we can write
\begin{eqnarray}
 \nabla_{X_{a}}\omega_{(n-1)}=\frac{1}{n}(div_{z}V)^{\ast}\tilde{X_{a}}\nonumber
\end{eqnarray}
for the KY $(n-1)$-form $\omega_{(n-1)}=^{\star}\tilde{V}$ (where
$V$ need not be a Yano vector). Equivalently, by the commutativity
of the Hodge map and covariant derivative, we have
\begin{eqnarray}
\nabla_{X_{a}}\tilde{V} =\frac{1}{n}(div_{z}V)\tilde{X_{a}} \;.
 \end{eqnarray}
On the other hand, for an arbitrary vector field $U$,
$\nabla_{X}\tilde{U}$ can be decomposed as follows:
\begin{eqnarray}
\nabla_{X}\tilde{U}=\frac{1}{2}i_{X}d\tilde{U}-
\frac{1}{n}\tilde{X}\delta\tilde{U}+\Gamma_X(U)\;,
\end{eqnarray}
(see \cite{Benn-Tucker} section 6.13 and \cite{Benn-Kress}) where
the 1-form $\Gamma_X(U)$ is defined by
\begin{eqnarray}
\Gamma_X(U)=\frac{1}{2}[(L_{U}g)-\frac{1}{n}tr(L_{U}g)g](X)\;.\nonumber
\end{eqnarray}
Thus, for each vector field $V$ corresponding to a KY $(n-1)$-form
we have, in view of (32), $i_{X}d\tilde{V}=-2\Gamma_X(V)$. That is,
$V$ is conformal if and only if $\tilde{V}$ is closed.

Let us return to equation (31): $\tilde{Y}$ is closed, obeys the
relation $\delta\tilde{Y}=-div_{z}Y$ and generates conformal
transformations $L_{Y}g=2\lambda g$ with the conformal weight
\begin{eqnarray}
\lambda=\frac{1}{2n}tr(L_{Y}g)=\frac{1}{2n}(L_{Y}g)(X^{a},
X_{a})\;.\nonumber
\end{eqnarray}
$Y$ is locally a gradient field for $\tilde{Y}$ is closed. From
(32), it also follows that
\begin{eqnarray}
\nabla_{Y}Y =\frac{1}{n}(div_{z}Y) Y \;,\nonumber
\end{eqnarray}
that is, the integral curves of $Y$ are pre-geodesics and may be
re-parameterized to become geodesics. If $Y$ is divergence-free,
then it is covariantly constant. We should note that in some
literature, a curve whose velocity vector $\dot{\gamma}$ satisfies
$\nabla_{\dot{\gamma}}\dot{\gamma}=f\dot{\gamma}$, for some function
$f$, is termed geodesic, which we here call pre-geodesic by adhering
to the nomenclature of \cite{Benn-Tucker,Benn-son}. For the Yano
vector we have $f=div_{z}Y/n$.

Substitution of (31) into (29) yields
$i_{X_{a}}F\wedge^{\ast}\tilde{Y}=0$, which in terms of components
can be rewritten as
\begin{eqnarray}
F_{ab}Y^{b}=0 \;.
 \end{eqnarray}
Note that condition (27) can also be rewritten as $F_{ab}K^{b}=0$,
where $K^{b}$'s are the components of the Killing vector $K$. Thus,
KY $(n-1)$-forms and 1-forms take part in the symmetry operator if
and only if the corresponding Yano and Killing vectors are contained
in the kernel of the matrix $F_{ab}$. These imply that if $F_{ab}$
is non-singular, which can happen only in even dimensions, no KY
$(n-1)$-form and 1-form can take part in the symmetry operator. Our
condition for KY 2-forms can also be stated as: a KY 2-form will
take part in the symmetry operator if and only if the corresponding
anti-symmetric matrix ($\omega_{ab}$) commutes with $F_{ab}$.

\subsection{Contractions with a Yano Vector}

Contractions of curvature 2-forms $R_{ab}$, Ricci 1-forms $P_{a}$
and conformal (Weyl) 2-forms $C_{ab}$, defined for $n>3$ by (below
${\cal R}$ denotes the scalar curvature)
\begin{eqnarray}
C_{ab}=R_{ab}-\frac{1}{n-2}(P_{a}\wedge e_{b}-P_{b}\wedge
e_{a}-\frac{\cal{R}}{n-1}e_{a}\wedge e_{b})\;,\nonumber
\end{eqnarray}
with a Yano vector enable one to reach decisive statements about the
global structures of the underlying space-times. The mentioned
contractions are found to be as follows:
\begin{eqnarray}
i_{Y}R_{ab}&=&\frac{1}{n-1}[(i_{Y}P_{a})e_{b}-(i_{Y}P_{b})e_{a}]\nonumber\\
Y_{b}i_{Y}P_{a}&=&Y_{a}i_{Y}P_{b}\;,\\
i_{Y}C_{ab}&=&\frac{1}{(n-1)(n-2)}[({\cal R} g_{ac}-P_{ac})e_{b}\nonumber\\
& &-({\cal R}
g_{bc}-P_{bc})e_{a}]Y^{c}+\frac{1}{n-2}(P_{a}Y_{b}-P_{b}Y_{a})\;,\nonumber
\end{eqnarray}
where $P_{ac}$ are the components of the Ricci tensor. The second
relation easily follows from a second application of $i_{Y}$ to the
first relation and the last one also follows from the definition of
$C_{ab}$ and the first relation of (35). A derivation of the first
relation is given in Appendix B. Equations (35) are generalizations
in the language of differential forms to an arbitrary number of
dimensions and signature of the tensorial relations first found by
McLenaghan and Spindel in the case of four dimensional Lorentzian
space-times.

If $P_{ab}={\cal R}g_{ab}/n$ such that ${\cal R}$ is constant, that
is in Einstein spaces, we have $i_{Y}C_{ab}=0$. In four dimensions,
the existence of a Yano vector on an Einstein space implies that the
space is conformally flat or of Petrov type N. In higher dimensions,
the equation $i_{Y}C_{ab}=0$ for nonzero $C_{ab}$ is necessary but
not sufficient condition for being Petrov type $N$. In such a case
the space-time can also be of Petrov type $II$ \cite{Coley}. When
$Y$ is non-null we can write the second equation of (35) as
$i_{Y}P_{a}=\lambda Y_{a}$, where $\lambda=Y^{b}i_{Y}P_{b}/g(Y,Y)$.
In such a case the Yano vector is an eigenvector of $P_{ab}$ with
eigenvalue $\lambda={\cal R}/n$. When $Y$ is null, it is also null
with respect to the Ricci tensor, in the sense that
$Y^{a}i_{Y}P_{a}=0$, in the directions of non-zero components of the
Yano vector. In other words, in such directions contraction of Ricci
forms with a null Yano vector projects it to an orthogonal
direction.

\section{Quadratic Geodesic Invariants and Constants of Motion}

The most important physical implication of the existence of KY-forms
is the fact that they provide first integrals of the geodesic
equation and may lead to quadratic functions of momenta that are
invariant also for the classical trajectories. As is shown in the
subsection B below, the condition (26) plays a crucial and unified
role in establishment of the second point.

\subsection{First Integrals of Geodesic Equations}

The easiest way to see the first point mentioned above is to
consider the relation $[\nabla_{X}, i_{Y}]=i_{\nabla_{X}Y}$ which
holds for two arbitrary vector fields $X, Y$. When $X$ and $Y$ are
equal to a velocity field $\dot{\gamma}$ of a geodesic, that is
$\nabla_{\dot{\gamma}}\dot{\gamma}=0$, the covariant and interior
derivatives commute and we obtain, in view of the defining relation
(11), $\nabla_{\dot{\gamma}} i_{\dot{\gamma}}\omega=0$ for any KY
$(p+1)$-form $\omega$. Therefore, the $p$-form
$\alpha=i_{\dot{\gamma}}\omega$ and hence its ``length"
$|\alpha|^{2}$ defined by $|\alpha|^{2}=g_{p}(\alpha,\alpha)$ remain
constant along the geodesic $\gamma$. Here $g_{p}$ is the compatible
metric induced by $g$ in the space of $p$-forms and we assume $g_0$
to simply multiply the two $0$-forms (see \cite{Benn-Tucker},
Chapter 1).

The constancy of $|\alpha|^{2}$ can be considered as a special case
of a more general fact \cite{Semmelman1}. To see this, let us
consider the symmetric bilinear form
\begin{eqnarray}
K^{(\beta)}(X,Y)=g_{p}(i_{X}\beta,i_{Y}\beta)=\epsilon^{\ast}(
i_{X}\beta\wedge^{\ast}i_{Y}\beta)\;,
\end{eqnarray}
defined, in terms of a $(p+1)$-form $\beta$ on the cartesian product
of the space of vector fields. Here $\epsilon$ denotes the sign of
the determinant of $g$. In the case of a KY $(p+1)$-form $\omega$,
one can easily verify the cyclic identity:
\begin{equation}
\nabla_{X}K^{(\omega)}(Y,Z)+\nabla_{Y}K^{(\omega)}(Z,X)+
\nabla_{Z}K^{(\omega)}(X,Y)=0\;.\nonumber
 \end{equation}
That is, the symmetrized covariant derivatives of $K^{(\omega)}$
vanish.  This shows that to any KY $(p+1)$-form $\omega$ is
associated a symmetric bilinear form $K^{(\omega)}$ which is the
Killing tensor generalizing the so-called Stackel-Killing tensor
that corresponds to a KY 2-form, first recognized by Penrose and
Floyd (the second tensor of (38) below). Since
\begin{equation}
\nabla_{X}[K^{(\omega)}(X,X)]=2K^{(\omega)}(\nabla_{X}X,X)\;,
\end{equation}
$K^{(\omega)}(\dot{\gamma},\dot{\gamma})$ is constant along the
geodesic $\gamma$.

For KY $1$-form $\omega_{(1)}=K_a e^{a}$, $2$-form
$\omega_{(2)}=2^{-1}\omega_{ab} e^{ab}$ and $n$-form
$\omega_{(n)}=kz$, where $k$ is a constant, the components of the
corresponding Killing tensor are found, respectively, to be
\begin{eqnarray}
K_aK_b\;,\;\;-\omega_{ac}\omega_{\;b}^{c}\;,\;\;\epsilon
k^{2}g_{ab}\;
\end{eqnarray}
In the case of KY $(n-1)$-form $^{\ast}\tilde{Y}$, the components of
$K^{(^{\ast}\tilde{Y})}$ can be written in terms of the Yano vector
$Y$ as
\begin{eqnarray} K^{(^{\ast}\tilde{Y})}_{ab}&=&
g_{n-2}(i_{X_a}{}^{\ast}\tilde{Y},i_{X_b}{}^{\ast}\tilde{Y})=
\epsilon g_{2}(\tilde{Y}\wedge \tilde{X}_a,\tilde{Y}\wedge
\tilde{X}_b)\;,\nonumber\\
&=&\epsilon [g(Y,Y) g_{ab}-Y_a Y_b]\;.
 \end{eqnarray}
This last Killing tensor, in particular, has remarkable properties
worth mentioning. By definition $K^{(^{\ast}\tilde{Y})}(Y,Y)=0$,
that is, Yano vectors are null with respect to the associated
Killing tensor. When $Y$ is non-null (that is, $g(Y,Y)\neq 0$ ),
$K^{(^{\ast}\tilde{Y})}$ has a one dimensional kernel spanned by
$Y$, and therefore it is singular. In that case every vector
orthogonal to $Y$ is an eigenvector with the eigenvalue $\epsilon
g(Y,Y)$ and the trace of $K^{(^{\ast}\tilde{Y})}$ is
$\epsilon(n-1)g(Y,Y)$. Moreover, the properly normalized Killing
tensor $K^{\prime}=K^{(^{\ast}\tilde{Y})}/\epsilon g(Y,Y)$ is an
idempotent projector $K^{\prime}_{ab}K^{\prime
b}_{\;\;c}=K^{\prime}_{ac}$ having rank $n-1$. When $Y$ is null we
have $K^{(^{\ast}\tilde{Y})}_{ab}=-\epsilon Y_a Y_b$ and in such a
case, it is a rank-1 nilpotent
($K^{(^{\ast}\tilde{Y})}_{ab}K^{(^{\ast}\tilde{Y})b}_{\:c}=0$)
projector, projecting vector fields to the direction of Yano vector.

\subsection{Constants of Motion for Classical Trajectories}

It is known that any rank-$p$ symmetric (covariant) Killing tensor
provides a degree-$p$ polynomial of velocity that is a geodesic
invariant \cite{Benn-son}. But, the second rank symmetric Killing
tensors that can be constructed from KY-forms, as in (36), have a
distinguished property of providing a quadratic invariant of
velocity along classical trajectories. In the remaining part of this
section we shall prove this statement in the most general setting of
this paper. To be precise, let us consider the quadratic function
\begin{equation}
f^{(\omega)}=K^{(\omega)}(u,u)=K^{(\omega)}_{ab}\,\:u^{a}u^{b}
\end{equation}
where $u=\dot{C}$ is the world velocity of a charged (charge assumed
to be unit) material particle obeying the classical equation of
motion
\begin{equation}
\nabla_{u}u=\frac{1}{m}\widetilde{i_{u}F}\;.
\end{equation}
Here, $u^{a}=dx^{a}/d\tau$ such that $x^{a}$'s are the local
coordinates of the world curve $C$ parameterized by the proper time
$\tau$ and $\nabla_{u}u$ represents the acceleration of the
particle. Hence
\begin{eqnarray}
\frac{d}{d\tau}(f^{(\omega)}\circ C)
&=&C_{\ast}(\partial_{\tau})f^{(\omega)}=
\nabla_{u}[K^{(\omega)}(u,u)]\nonumber\\
&=&2K^{(\omega)}(\nabla_{u}u,u)\;,
\end{eqnarray}
and in view of (41)
\begin{eqnarray}
\frac{d}{d\tau}(f^{(\omega)}\circ C)=
\frac{2}{m}u^{a}u^{b}F_{ac}K^{(\omega)c}_{\;b}.
\end{eqnarray}

What we are going to prove is the constancy of $f^{(\omega)}$ along
the classical trajectory determined by (41). Before doing that, it
would be illuminating to examine first some special cases. One can
easily verify that the right hand side of (43) vanishes if the
components of the Killing tensor given by (38) are used, provided
that KY-forms employed in defining $f^{(\omega)}$ satisfy the
condition (26). In the case of $K^{(^{\ast}\tilde{Y})}_{ab}$ given
by (39), the contraction with the Maxwell field is, in view of (34)
\begin{equation}
F_{ac}K^{(\omega)c}_{\;b}=\epsilon g(Y,Y)F_{ab} \;,
\end{equation}
which is obviously anti-symmetric and also makes the right hand side
of (43) vanish. So, as a particular result, all quadratic functions
constructed from KY $p$-forms for $p=1,2, n-1, n$ are constant along
the classical trajectory. These exhaust all possible cases in a four
dimensional space-time and, to the best of our knowledge, these are
all that can be found in the related literature.

We shall now prove that $f^{(\omega)}$ is a constant of motion for
the classical trajectories determined by (41) for all dimensions and
signatures as well as for all KY $p$-forms $\omega_{(p)}$ obeying
the symmetry condition (26). For the proof, we first take the Hodge
dual of both sides of the equation (43) and write
\begin{eqnarray}
\frac{d}{d\tau}(f^{(\omega)}\circ C)^{\ast}1=
\frac{2}{m}u^{a}u^{b}I_{ab}\;,
\end{eqnarray}
where the $n$-form $I_{ab}$ is defined as
\begin{equation}
I_{ab}=^{\ast}(F_{ac}K^{(\omega)c}_{\;b})=i_{X_c}i_{X_a}Fi_{X^c}\omega\wedge^{\ast}
i_{X_b}\omega\;.
\end{equation}
In passing to the second equality of equation (46) we have used
(36). Thus, $f^{(\omega)}$ is constant along the world line
$C(\tau)$ if and only if the right hand side of the equation (45)
vanishes. Evidently, the anti-symmetry condition $I_{ab}=-I_{ba}$ is
sufficient (and can also be shown to be necessary) for the right
hand side of (45) to vanish. To show the anti-symmetry of $I_{ab}$
we rewrite it as follows:
\begin{eqnarray}
I_{ab}&=&-[i_{X_a}(i_{X_c}F\wedge i_{X^c}\omega)+i_{X_c}F\wedge
i_{X_a}i_{X^c}\omega]\wedge^{\ast}i_{X_b}\omega\;,\nonumber\\
&=&i_{X_c}F\wedge i_{X^c}(i_{X_a}\omega\wedge^{\ast}i_{X_b}\omega)-
i_{X_a}\omega\wedge i_{X_c}F\wedge i_{X^c}\;^{\ast}i_{X_b}\omega\;,\\
&=&-i_{X_a}\omega\wedge i_{X_c}F\wedge^{\ast}(i_{X_b}\omega\wedge
e^{c})\;.\nonumber
\end{eqnarray}
The first term in the square bracket of the first line vanishes
because of condition (26), and the first term of the second line
vanishes since it can be written as an interior derivative of a
$(n+1)$-form. In the third line of (47) we have made use of the
identity $i_{X^c}\;^{\ast}i_{X_b}\omega=^{\ast}(i_{X_b}\omega\wedge
e^{c})$ which was also used in the previous section. In view of
$i_{X_c}F=F_{ck}e^{k}$ and of another Hodge identity
$\alpha\wedge^{\ast}\beta=\beta\wedge^{\ast}\alpha$ which holds for
any two $p$-forms $\alpha$ and $\beta$, we obtain from (47)
\begin{eqnarray}
I_{ab}&=&-i_{X_b}\omega\wedge
F_{ck}e^{c}\wedge^{\ast}(i_{X_a}\omega\wedge
e^{k})\nonumber\\
&=&i_{X_b}\omega\wedge i_{X_k}F\wedge^{\ast}(i_{X_a}\omega\wedge
e^{k})\;.\nonumber
\end{eqnarray}
By comparing this with third line of (47) we obtain $I_{ab}=-I_{ab}$
which was what to be demonstrated.

An immediate corollary of the constancy of $f^{(\omega)}$ is, by
virtue of (42), the relation $K^{(\omega)}(\nabla_{u}u,u)=0$ which
can be interpreted as follows. The world velocity and acceleration
of a charged material particle are perpendicular to each other, not
only with respect to the metric, but also with respect to the
symmetric Killing tensors associated with each KY-form satisfying
the condition (26).

\section{Summary and Conclusion}

In this study, the most general first-order linear symmetry
operators of the Dirac equation including interaction with Maxwell
field in curved background of arbitrary $n$-dimension and of
signature are specified. We have shown that all coefficients forms
$\omega^{a}$'s of the symmetry operator $L=2\omega^{a}
S_{X_a}+\Omega$ are given, in terms of an inhomogeneous KY-form
$\omega$, by $\omega^{a}=i_{X^{a}}\omega$. The components of
$\Omega$ are explicitly calculated and are given by the equations
(12), (13) and (28). They depend on the exterior derivative of
KY-forms and their contraction with the potential field $A$. We have
also found a unified, gauge invariant dynamical symmetry condition
which states that among all the KY-forms of underlying curved
background only those which Clifford commute with the Maxwell field
can take part in $L$. When $\omega^{a}$ and $\Omega$ are even, $L$
itself, but when they are odd and $n$ is even $Lz$ is a first order
symmetry operator which Clifford commutes with the Dirac equation
and hence maps a solution to another.

The special cases of the dynamical symmetry condition are also
discussed as they may provide valuable insights in applications. In
particular, the KY $(n-1)$-forms and 1-forms take part in the
symmetry operator if and only if the corresponding Yano and Killing
vectors belong to the kernel of the anti-symmetric matrix $F_{ab}$
corresponding to the components of the Maxwell field. These imply
that if $F_{ab}$ is non-singular, which can happen only in even
dimensions, no KY $(n-1)$-forms and 1-forms can take part in the
symmetry operator. Implications of the existence of Yano vectors in
specifying the global structure of the curved background are also
discussed. Finally it has been proved that, for all KY-forms obeying
the dynamical symmetry condition, there exists a quadratic function
of velocity (defined by the equation (40)) which is a constant of
motion for the classical motion in an arbitrary dimension and
signature.

All of these results are expected to provide a unified framework for
symmetry analysis and to serve as a firm base in studying symmetry
algebras and related conserved quantities of the Dirac equation for
a given specific curved background and force field.

\begin{acknowledgments}
We are grateful to anonymous referees for their useful comments.
This work was supported in part by the Scientific and Technological
Research Council of Turkey (T\"{U}B\.{I}TAK).
\end{acknowledgments}

\begin{appendix}

\section{Consistency Conditions}

By direct calculations we obtain
 \begin{eqnarray}
 [\displaystyle{\not}S, \omega]=
 2\omega^{a}S_{X_a}+\displaystyle{\not}d \omega=L+\varphi\;,\nonumber
\end{eqnarray}
where $\omega^{a}=i_{X^{a}}\omega$. It proves convenient to take $L$
as $L=[\displaystyle{\not}S, \omega]-\varphi$, where
$\Omega=\displaystyle{\not}d \omega-\varphi$. Using the fact that
 \begin{eqnarray}
 [\displaystyle{\not}S, [\displaystyle{\not}S, \omega]]=
 [\displaystyle{\not}S^{2}, \omega]_{-}\;,\nonumber
 \end{eqnarray}
 the symmetry condition (3) transforms to
 \begin{eqnarray}
 [\displaystyle{\not}S^{2}, \omega]_{-}=
 [\displaystyle{\not}S, \varphi]-{\rm{i}}[A,L]\;.
  \end{eqnarray}
 ${\cal R}$ being the scalar curvature of the spinor connection,
 $\displaystyle{\not}S^{2}$ acts on the spinor fields as (see
 \cite{Benn-Tucker} chapter 10)
 \begin{eqnarray}
 \displaystyle{\not}S^{2}=S^{2}(X_{a},X^{a})-\frac{1}{4}{\cal R}\;.
 \end{eqnarray}
 Therefore  on substituting the relations
\begin{eqnarray}
[\displaystyle{\not}S^{2}, \omega]_{-}&=&
[S^{2}(X_{a},X^{a}),\omega]_{-}\nonumber\\
&=& 2\nabla_{X^{a}}\omega S_{X_{a}}+\nabla^{2}(X_{a},X^{a})
\omega\;,\nonumber\\
\;[\displaystyle{\not}S, \varphi]&=&
2i_{X^{a}}\varphi S_{X_a}+\displaystyle{\not}d\varphi\;,\\
\;[A, L]&=& 2[A,i_{X^{a}}\omega]S_{X_{a}}-
2(i_{X^{a}}\omega)^{\eta}\nabla_{X_{a}}A +[A,\Omega],\nonumber
\end{eqnarray}
into equation (A1) and then on equating the coefficients of equal
power of derivatives, we obtain the consistency conditions (6) and
(7) of the main text.

\section{Contraction of Curvatures with a Yano vector}

By differentiating the KY equation (11), the action of the Hessian
(see the remark following equation (4)) on a KY $p$-form
$\omega_{(p)}$ is found to be
\begin{equation}
\nabla^2(X_a,X_b)\omega_{(p)}=\frac{1}{p+1}i_{X_b}\nabla_{X_a}d\omega_{(p)}.
\end{equation}
Since  the difference of $\nabla^2(X_a,X_b)$ and $\nabla^2(X_b,X_a)$
is the curvature operator $R(X_a,X_b)$, from (B1) we obtain
\begin{equation}
R(X_a,X_b)\omega_{(p)}=\frac{1}{p+1}(i_{X_b}\nabla_{X_a}-i_{X_a}\nabla_{X_b})d\omega_{(p)}.
\end{equation}
On the other hand, the action of the curvature operator on any form
$\alpha$ is known to be (see \cite{Benn-Tucker} equation (8.1.11)
and \cite{Benn-Kress})
\begin{equation}
R(X_a,X_b)\alpha=-i_{X^k}R_{ab}\wedge i_{X_k}\alpha\;.
\end{equation}
Note that the action on the one form $\tilde{X}$ simply is
$-i_{X}R_{ab}$. Using (B3) in (B2) and then by multiplying the
result with $e^a\wedge$, we obtain
\begin{equation}
\nabla_{X_a}d\omega_{(p)}=\frac{p+1}{p}R^b_{\;\;a}\wedge
i_{X_b}\omega_{(p)}.
\end{equation}
Using this in (B2) for $\omega_{(n-1)}=^{\ast}\tilde{Y}$ we find
\begin{equation}
R(X_a,X_b)^{\ast}\tilde{Y}=\frac{(-1)^{n}}{n-1}[
i_{X_b}{}^{\ast}\tilde{Y}\wedge P_a-i_{X_a}{}^{\ast}\tilde{Y}\wedge
P_b]
\end{equation}
where we have used the contracted Bianchi identity
$i_{X_b}P_a=i_{X_a}P_b$. Let us first note that
\begin{eqnarray}
R(X_a,X_b)^{\ast\ast}\tilde{Y}&=&\epsilon(-1)^{n}i_{Y}R_{ab}\;,\\
^{\ast}[^{\ast}(\tilde{Y}\wedge e_b)\wedge P_a]&=&
i_{\tilde{P_a}}{}^{\ast\ast}(\tilde{Y}\wedge e_b)\;,\nonumber\\
&=&\epsilon [(i_{Y}P_a)e_b-\tilde{Y}i_{X_b}P_a]\;,
\end{eqnarray}
where $\epsilon$ is the sign of $\det g$. In deriving the above
relations, we have made use of (B3) and of the following identities:
\begin{eqnarray}
^{\ast\ast}\alpha_{(p)}=\epsilon(-1)^{p(n-p)}\alpha_{(p)}\;,\quad
i_{X_a}{}^{\ast}\tilde{Y}=^{\ast}(\tilde{Y}\wedge e_a)\;.
\end{eqnarray}
If we now take the Hodge dual of both sides of (B5) and use (B6),
(B7) we obtain the first relation of (35) of the main text:
$i_{Y}R_{ab}=(n-1)^{-1}[(i_{Y}P_{a})e_{b}-(i_{Y}P_{b})e_{a}]$.

As an aside, it is worth mentioning that if (B4) is substituted into
(B1), we obtain
\begin{equation}
\nabla^2(X_a,X_b)\omega_{(p)}=\frac{1}{p}
i_{X_{b}}(R^{c}_{\;a}\wedge i_{X_{c}}\omega_{(p)}),
\end{equation}
which shows that the second covariant derivative of any KY $p$-form
is determined by curvature characteristics and the form itself. This
implies that the value of a KY $p$-form at any point is entirely
determined by the values of the form itself and its first covariant
derivatives at the same point. These remarks can be used to
determine the upper bound for the numbers of linearly independent KY
$p$-forms to be the Binomial number $C(n+1,p+1)$
\cite{Kastor,ozumav1}.
\end{appendix}

\end{document}